\documentstyle[12pt]{article}
\newcommand{\be}{\begin{equation}}
\newcommand{\ee}{\end{equation}}

\newcommand{\go}{\omega}
\newcommand{\half}{\frac{1}{2}}
\newcommand{\ga}{\alpha}

\newcommand{\gb}{\beta}
\newcommand{\gee}{\epsilon}
\newcommand{\ggg}{\gamma}

\newcommand{\bee}{\begin{eqnarray}}
\newcommand{\eee}{\end{eqnarray}}
\begin{document}

\begin{flushright}
\vspace{-1mm}
 FIAN/TD/21--96\\
\vspace{-1mm}
{July 1996}\\
\end{flushright}\vspace{2cm}

\begin{center}
{\large\bf
HIGHER-SPIN-MATTER INTERACTIONS IN 2+1 DIMENSIONS}
\vglue 0.6  true cm
{\bf M.A.~VASILIEV}
\vglue 0.3  true cm

I.E.Tamm Department of Theoretical Physics, Lebedev Physical Institute,\\
Leninsky prospect 53, 117924, Moscow, Russia
\vskip2cm
Invited talk at the $10^{th}$ International Conference Problems of
 Quantum Field Theory, Alushta (Cremea - Ukrain), May 13-18, 1996.
\vskip2cm
\end{center}
\vskip0.2cm
\begin{abstract}
We describe a model of massive matter fields
interacting through higher-spin gauge
fields in 2+1  dimensional space-time.
The two main conclusions are that the
parameter of mass $M$
appears as a module characterizing an appropriate vacuum
solution of the full non-linear model and
that $M$ affects a structure of a global vacuum
higher-spin symmetry which
leaves invariant the chosen vacuum solution.
\end{abstract}

\vskip0.6cm

\newpage
\section{Introduction}
Higher-spin (HS) symmetries are (infinite-dimensional) symmetries
generated by conserved currents which contain higher derivatives
of dynamical fields.
HS gauge theories are
most symmetric gauge theories which can
underlie a fundamental theory of unified interactions.
In 3+1 space-time dimensions HS gauge fields
 describe massless fields of arbitrarily
high spin which have their own degrees of freedom \cite{F}.
In lower dimensions
HS gauge fields do not propagate rather mediating interactions
of matter
sources analogously to the case of the gravitational field in
2+1 dimensions \cite{T,W}. Analysis of HS interactions
of relatively simple
lower dimensional models is useful since
it sheds some light on
general properties of HS models. In this report
which is based on a recent work completed with S.~Prokushkin
we focus on the
specificities of the HS interactions for the case of
massive matter fields.

Originally it was conjectured by Blencowe \cite{bl} that
a 3d HS algebra is the direct sum
of two Heisenberg-Weyl algebras (more precisely, of
their Lie supercommutator
superalgebras). More generally, 3d HS gauge fields $A$ may
correspond to any algebra $g$ which contains  the anti-de Sitter (AdS) algebra
$o(2,2)\sim sp(2)\oplus sp(2)$ as a subalgebra and admits a non-degenerate
invariant bilinear form which allows one to write
the Chern-Simons action for the pure gauge HS system,
$
S=\int_{M_3}str (A\wedge dA +\frac{2}{3}A\wedge A\wedge A )\,.
$
In \cite{BBS,H,Q} it was shown that there exists a one-parametric class
of infinite-dimensional algebras which we denote $ hs(2;\nu )$
($\nu$ is an arbitrary real parameter), all containing $sp(2)$ as a
subalgebra. This allows one to define a class of HS algebras
$
g=hs(2;\nu )\oplus hs(2;\nu ).
$
The supertrace operation
was defined in \cite{Q} where also the following
useful realization of
(supersymmetric extension of)
$ hs(2;\nu )$ was given.

Consider an associative algebra
$Aq(2,\nu )$ with a general element of the form
\begin{equation}
\label{sel}
f(q,K )=
\sum_{n=0}^{\infty}\sum_{A=0,1}
\frac{1}{2i\,n!}
 f^{A\,\alpha_1\ldots\alpha_n}(K)^A
 q_{\alpha_1}\ldots q_{\alpha_n}\, \,,
\end{equation}
under condition that the coefficients
$ f^{A\,\alpha_1\ldots\alpha_n}$ are symmetric with respect to
 the indices
$\ga_j =1,2$ while the generating elements $q_\ga$ satisfy the
 relations
\be
\label{modosc}
[q_\ga ,q_\gb ]=2i \gee_{\ga\gb} (1+\nu K)\,;\qquad Kq_\ga =-q_\ga K\,;
\qquad K^2 =1\,;\qquad \gee_{ 12} =1\,.
\ee
The two important properties of this algebra are that $(i)$
it admits the
following unique supertrace operation \cite{Q}
$str (f) =f^0 -\nu f^1$, such that
$str(fg)$ = $(-1)^{\pi_f \pi_g} str(gf)$, $\forall f,g $
having a definite parity, $f(-q,K)
=(-1)^{\pi_f}f(q,K)$ (i.e. $str (1)=1\,,\quad str(K) =-\nu$ and all higher
monomials of $q_\ga$ in (\ref{sel}) do not contribute under the supertrace)
and $(ii)$ that for all $\nu$ the bilinears
$T_{\ga\gb}=\frac{1}{4i}\{q_\ga \,,q_\gb \}$
have $sp(2)$
commutation relations and rotate $q_\ga$ as a $sp(2)$ vector
\be
[T_{\ga \gb}, T_{\ggg \eta}]=((\gee_{\ga\ggg}T_{\gb\eta}
+\ga\leftrightarrow\gb)
+\ggg\leftrightarrow\eta )\,;
\quad
[T_{\ga \gb}, q_{\ggg }]=\gee_{\ga\ggg}q_{\gb}
+\ga\leftrightarrow\gb\,.
\ee

To describe a
doubling of the elementary algebras in
$g=hs(2;\nu )\oplus hs(2;\nu )$
it is convenient to
introduce an additional central involutive
 generating element $\psi$:
$[\psi ,q_\ga ]=0$, $[\psi ,K]=0$,
 $\psi^2 =1$. The two simple subalgebras
of $g$ are singled out by the projection operators
$P_\pm$ = $\half (1\pm \psi )$.

The full set of HS gauge fields in 2+1 dimensions
thus is
\be
\label{gau}
A(q,K,\psi |x)=dx^\nu \sum_{n=0}^\infty\sum_{B=0,1}
\frac{1}{n\,!} (\go_\nu^{B\ga_1 \ldots \ga_n }(x)+\psi
h_\nu^{B\ga_1 \ldots \ga_n }(x))\,(K)^B\,
q_{\ga_1}\ldots q_{\ga_n}\,.
\ee
The field strengths and gauge transformation laws are defined
in the usual way
$$
R(q,K,\psi |x)=d
A(q,K,\psi |x)
+A(q,K,\psi |x)\wedge
A(q,K,\psi |x)\,,\quad d=dx^\nu \frac{\partial}{\partial x^\nu }\,;
$$
$$
\delta A(q,K,\psi |x)=d
\gee(q,K,\psi |x)
+[A(q,K,\psi |x)\wedge
\gee (q,K,\psi |x)].
$$
The gravitational fields
$A^{gr} =\half (\go^{\ga\gb} +h^{\ga\gb}\psi )q_\ga q_\gb$
take values in the subalgebra $sp(2)\oplus sp(2)$.
The pure gauge HS action has the Chern-Simons
form. It reduces to the Witten gravity action
\cite{W} in the
spin 2 sector and to the Blencowe's HS action \cite{bl}
in the case of $\nu =0$.
An important question is then how to introduce
interactions of HS gauge fields with propagating matter fields.

\section{Unfolded equations}
In this report we will show how this problem can be solved at the
level of equations of motion using an approach
 which we call ``unfolded formulation" \cite{un}.
It consists of rewriting dynamical equations in a form of certain
 zero-curvature conditions and covariant constantness
 conditions
\be
\label{0cur}
d\go =\go\wedge\go\,,\qquad
dB^A =\go^i t_i \, {}^A{}_B B^B\,,
\ee
supplemented with some gauge invariant constraints
\be
\label{con}
\chi (B) =0 \,
\ee
which do not contain
space-time derivatives.
Here
 $\go (x)=dx^\nu \go_\nu^i (x) T_i $ is a gauge field
taking values in some Lie superalgebra $l$ ($T_i \in l)$,
and $B^A (x)$ is a set of 0-forms  which take values in
a representation space of
some representation $(t_i ){}^B {}_A$ of $l$.

An interesting property of this form of
equations is that their
dynamical content
is hidden  in the constraints (\ref{con}).
Indeed, locally one can integrate out explicitly
(\ref{0cur}) as
$\go=d g (x)g^{-1} (x)$ and
$B(x)=t_{g(x)} (B_0 )$
where $g(x)$ is an arbitrary invertible element
while
$B_0$ is an arbitrary $x$ - independent
representation element and
$t_{g(x)}$ is the exponential of the representation $t$ of $l$.
Since the constraints $\chi (B)$ are gauge invariant one is left with the
only condition
$\label{con0}
\chi (B_0 )=0\,.$
Let $g(x_0 )=I$ for some point of space-time  $x_0$.
Then $B_0$=$B(x_0 )$. One can wonder
how any restrictions on
values of some 0-forms in a fixed point of space-time can lead
to a non-trivial dynamics. This is possible
if the set of 0-forms $B$ is reach enough to describe
all space-time derivatives of dynamical fields
while the constraints (\ref{con}) just impose all
restrictions on the space-time derivatives required by the
dynamical equations under consideration.
Given solution
of (\ref{con}) one knows all derivatives of the dynamical fields
 compatible with the field equations and can therefore reconstruct
these fields by analyticity in some neighborhood of $x_0$.
The specificity of the HS dynamics
which makes such an approach adequate is that HS
 symmetries mix all orders of derivatives which therefore
 have to be contained in the
 representations $t$  of
 HS symmetries.

 To illustrate this point let us consider an example
 of a scalar field $\phi$ obeying the massless Klein-Gordon
 equation $\Box\phi=0$ in a flat  space-time
 of an arbitrary dimension $d$.
 Here $l$ is identified with
the Poincare algebra $iso(d-1,1)$ which gives rise to the gauge fields
$
\go_\nu =(h_\nu{}^a ,\go_\nu {}^{ab} )
\quad$
($a,b =0-(d-1)$).
The zero curvature conditions of $iso(d-1,1)$,
$
R_{\nu\mu}{}^a =0,$ $ R_{\nu\mu}{}^{ab}=0
$,
imply that the vierbein $h_\nu {}^a$
and Lorentz connection $\go_\nu {}^{ab}$ describe the flat geometry.
Fixing the local Poincare gauge transformations one can set
$
h_\nu {}^a =\delta _\nu^a \,,\quad \go_\nu{}^{ab}=0.
$

To describe dynamics of a spin zero massless field
$\phi (x)$ let us introduce an
infinite collection of 0-forms $\phi_{a_1\ldots a_n}(x)$
which are totally symmetric traceless tensors
$\eta^{bc}\phi_{bca_3\ldots a_n}=0\,,$
where $\eta^{bc}$ is the flat Minkowski metrics.
The ``unfolded" version of the Klein-Gordon equation
has a form of the following infinite chain of equations
\be
\label{un0}
\partial_\nu \phi_{a_1\ldots a_n }(x) =h_\nu {}^b
\phi_{a_1 \ldots a_n b}(x)\,,
\ee
where we have replaced the Lorentz covariant
derivative by the ordinary flat derivative $\partial_\nu$ using the
gauge condition $\go _{\nu ,ab}=0$.
The tracelessness condition for $\phi$  is a specific realization of the
constraints (\ref{con}) while the system of equations
(\ref{un0}) is a particular example of the equations (\ref{0cur}).
It is easy to see that this system is formally
consistent, i.e. $\partial_\mu$ differentiation of (\ref{un0})
does not lead to new conditions after antisymmetrization
$\nu\leftrightarrow\mu$. This property is equivalent to the fact that
the set of zero forms $\phi_{a_1 \ldots a_n}$ spans some
representation of the Poincare algebra.

To show that the system (\ref{un0}) is equivalent to the
free massless field equation $\Box \phi (x)=0$ let us identify the
scalar field $\phi (x)$ with the $n=0$ member of the tower
of 0-forms $\phi_{a_1 \ldots a_n}$.
Then the first two equations
(\ref{un0}) read $\partial_\nu \phi =\phi_\nu$ and
$
\partial_\nu \phi_\mu= \phi_{\mu\nu}$, respectively.
The first one tells us that
$\phi_\nu$ is a first derivative of $\phi$.
The second one implies that
$\phi_{\nu\mu}$ is a second derivative of $\phi$. However, because of the
tracelessness condition for $\phi_{\nu\mu}$ it imposes the Klein-Gordon
equation
$\Box\phi =0$.
It is easy to see that all other equations in (\ref{un0}) express highest
tensors in terms of the higher-order derivatives
$
\phi_{\nu_1 \ldots \nu_n}= \partial_{\nu_1}\ldots\partial_{\nu_n}\phi
$
and impose no additional conditions on $\phi$. The tracelessness conditions
are all satisfied once the Klein-Gordon equation is true.

\section{Free fields in 2+1 AdS space}
Let us now confine ourselves to the 2+1 dimensional
case and generalize the above analysis of the scalar
field dynamics to the AdS geometry. The gauge fields
of the AdS algebra $o(2,2)\sim sp(2)\oplus sp(2)$
are identified with the gravitational
fields $A_\nu =(\lambda h_{\nu \,,\ga\gb}\,;\go_{\nu \,,\ga\gb} )$.
The zero-curvature conditions $R_{\nu\mu}=0$ for the AdS algebra
in its orthogonal realization take a form
\be
\label{00cur}
R_{\nu\mu\,,ab}=\lambda^2 (h_{\nu a}h_{\mu b} -h_{\nu b}h_{\mu a})\,,\qquad
R_{\nu\mu\,,a}=0 \qquad (\nu\,,\mu\ldots ; a\,,b\ldots =0-2)\,,
\ee
where $R_{\nu\mu\,,ab}$ and $R_{\nu\mu\,,a}$ are the
Riemann and torsion tensors, respectively. {}From (\ref{00cur})
one concludes that the zero curvature
equations for the algebra $o(2,2)$ on a 3d
manifold does indeed describe the AdS space provided that
$h_\nu {}^a$ is identified with a dreibein and is invertible.

It is an important property of the 3d geometry
 that
one can resolve the tracelessness conditions for
symmetric tensor fields by using the formalism of two-component
spinors: a totally symmetric traceless tensor $\phi _{\mu_1 \ldots \mu_n}$
is equivalent to a totally symmetric multispinor
$C_{\ga_1 \ldots \ga_{2n}}$. Let us now address the question
what is a general form of the
 equations analogous to (\ref{un0})
such that their   integrability conditions reduce to (\ref{00cur}).
The result is  that these are equations of the
form \cite{un}
\be
\label{unads}
D C_{\ga_1 \ldots \ga_{2n}} =h^{\gb\gamma}C_{\ga_1 \ldots\ga_{2n} \gb\gamma}
+2n(2n-1) e(2n,\lambda,M)h_{\{ \ga_1\ga_2} C_{\ga_3 \ldots \ga_{2n}\}_\ga}\,,
\ee
where $D$ is the Lorentz covariant derivative,
$D B_\ga =dB_\ga +\go_\ga{}^\gb B_\gb$, and
$e(l ,\lambda ,M)=\frac{1}{4}\lambda^2 -\half\frac{M^2}{l^2 -1}$,
$( l\geq 2). $
$M$ is an arbitrary parameter.
One can see that this freedom in
$M$ is just the freedom of the relativistic
field equations in the
parameter of mass.

Thus the equations (\ref{unads}) describe
a scalar field of an
arbitrary mass
in 2+1 dimensions. Now let us show how these equations can be
generated
with the
aid of the generalized oscillators (\ref{modosc}). To this end we
introduce the generating function
\be
\label{gf}
C(q_\ga ,K|x) =\sum_{A=0,1}\sum^\infty_{n=0} \frac{1}{n!}C_{\ga_1 \ldots \ga_n
}(x)(K)^A q^{\ga_1}\ldots q^{\ga_n}\,.
\ee
The relevant equations acquire then
the following simple form
\be
\label{coeq} D^L C(q_\ga ,K|x) =\{ h^{\ga\gb}
q_\ga q_\gb , C(q,K |x)\}\,
\ee
(from now on we use the dimensionless units
with a unit AdS radius, $\lambda =1$).  To see that the integrability
conditions
for (\ref{coeq})  reduce to the zero-curvature conditions for $sp(2)\oplus
sp(2)$ one observes that there is an automorphism of the AdS algebra which
changes a
sign of the AdS translations. This automorphism allows
one to introduce a ``twisted representation" of the AdS algebra
with the anticommutator (instead of the commutator) in the translational
part of the AdS algebra. This twisted representation just leads to the
covariant constantness equations (\ref{coeq}).

Since the part which depends on the background gauge fields
in (\ref{coeq}) only involves even combinations of the oscillators
$q_\ga$ the full system of equations decomposes into four independent
subsystems which can be singled out by virtue of the
projection operators
$P_\pm =\half (1\pm K)$ either in the boson or in the fermion sectors
(even (odd) functions
$C(q_\ga ,K|x)$ of
$q_\ga$
describe bosons (fermions)).
The explicit calculation which involves some reorderings of $q_\ga$
and rescalings of fields
then shows that the irreducible boson subsystems
 projected out by $P_\pm$ indeed reduce to the
equations of motion of the form (\ref{unads}) for
a massive scalar field
of mass $M^2 =\half\nu (\nu \mp 2)$. Remarkably, the same equations in the
fermion sector describe spin $\half$
fermion fields of the mass $M^2=\half\nu^2$.

An important achievement of the reformulation of the free field
equations in the form (\ref{coeq}) is that this form suggests
that the global HS symmetry algebra realized on the matter fields
of mass $M(\nu )$ is
$g=hs(2;\nu )\oplus hs(2;\nu )$ with the gauge fields (\ref{gau}).
However to simplify the formulation it is convenient to introduce  two
Clifford variables
$\psi_{1,2}$ ($\{\psi_i ,\psi_j \}=\delta_{ij})$
instead of $\psi$.
One then introduces the full set of HS gauge fields as
$W_\nu (q_\ga ,K,\psi_{1,2} |x)$ and realizes
the gravitational fields as
\be
\label{gr}
W_\nu^{gr} =\half (\go_\nu{}^{\ga\gb}+
h_\nu{}^{\ga\beta}\psi_1 )q_\ga q_\beta\,.
\ee
The generating function for
0-forms is
\be
\label{maux}
C( q,K,\psi_{1,2} |x)=
C^{mat} ( q,K,\psi_1 |x)\psi_2 +
C^{aux} ( q,K,\psi_1 |x)\,.
\ee

Now let us consider the zero curvature equations
$0=R\equiv dW (q,K,\psi |x )+
W (q,K,\psi |x)\wedge
W (q,K,\psi |x)
$
along with the covariant constantness conditions in the adjoint
representation of the HS algebra
\be
0={\cal D}C\equiv dC (q,K,\psi |x)+
W (q,K,\psi |x)C (q,K,\psi |x)-
C (q,K,\psi |x)W (q,K,\psi |x)\,.
\ee
Due to the factor of $\psi_2$ in front of $C^{mat}$
the equations for $C^{mat}$ turn out to be
equivalent to the equations (\ref{coeq}) in the gauge in which
only the gravitational part (\ref{gr}) of the set of HS
gauge fields is non-vanishing. The fields $C^{aux}$
can be shown \cite{un} to be of a topological type
so that each irreducible subsystem in this sector can describe
at most a finite number of degrees of freedom and trivializes in
a topologically trivial situation. Thus the effect of introducing
a second Clifford element  consists of addition of some topological
fields.

\section{Non-linear dynamics} To describe non-linear HS dynamics of
matter fields in 2+1 dimensions we start with a system of
equations which is very close to that introduced in \cite{eq}
for a particular case of massless matter fields. We
introduce three types of the generating functions
$dx^\nu W_\nu (z_\ga ,y_\gb ,K,\psi_i |x)$,
$ s_\gamma (z_\ga ,y_\gb ,K,\psi_i |x)$ and
$ B (z_\ga ,y_\gb ,K,\psi_i |x)$ which depend on the
space-time variables $x^\mu$ and auxiliary variables
 $(z_\ga ,y_\gb ,K,\psi_i )$ such that
 the two Clifford elements $\psi_i$
commute to all other variables, while the bosonic spinor
variables $z_\ga$ and $y_\gb$ commute to each other but anticommute
with $K$, $\{K,z_\ga \}= \{K,y_\ga \}=0,\quad K^2=1$.
Their physical content
is as follows:
$dx^\nu W_\nu $ is the generating
function for HS gauge fields,
$ B $ contains physical matter degrees
of freedom along with some auxiliary variables,
and $ s_\gamma $ is entirely
auxiliary variable which  allows one to formulate the
full system of equations in a compact form. This formulation
is based on the following star-product law which endows the space
of functions $f(z,y)$ with a structure of associative
algebra
\begin{equation}
\label{star}
(f*g)(z,y)=(2\pi )^{-2}\int d^2 ud^2 v\, f(z+u,y+u)\,g(z-v,y+v)\,\exp
i(u_\alpha v^\alpha )\,\,.
\end{equation}
This product law provides a
particular symbol realization of the Heisenberg--Weyl algebra,
$ [y_\ga ,y_\gb ]_* =-[z_\ga ,z_\gb ]_* =2i
\epsilon_{\ga\gb}\,$.

The full system of equations has the form:
\be
\label{zercur}
dW+W*\wedge W=0\,,\qquad
ds_\ga +\tilde{W}* s_\ga -s_\ga * W=0\,,\qquad
dB +{W}* B -B * W=0\,,
\ee
and
\be
\label{cons}
\tilde{s}_\ga * s_\gb
-\tilde{s}_\gb * s_\ga=-2i\gee_{\ga\gb} (1+\kappa * B)\,,\qquad
\tilde{B}* s_\ga -s_\ga * B=0\,,
\ee
where $\kappa =K exp\,i(z_\ga y^\ga )$ is a central element of the
algebra which has vanishing star commutators with $y_\ga$, $z_\ga$,
$K$ and $\psi_i$, while  $\tilde{a}(z ,y ,K, \psi_i |x)
=a(z ,y ,-K, \psi_i |x)$ $\forall a$.

The system of equations (\ref{zercur}),(\ref{cons})
is explicitly invariant
under the general coordinate  transformations
and the
HS gauge transformations
$\delta W=d\gee +W*\gee -\gee* W,$
$\delta B=B*\gee -\gee* B$ and
$\delta s_\ga = s_\ga *\gee -\tilde{\gee}* s_\ga.$
To elucidate its physical content
one has to analyze this system perturbatively near some
vacuum solution. In the massless case the appropriate vacuum solution
\cite{eq} is
$B_0 =0$, $ s_0 {}_\ga =z_\ga$ and $ W_0 =\go (y,K,\psi_{1,2} )$
with the vacuum gauge field $\go$ satisfying the zero curvature
condition $d\go +\go*\wedge\go=0$.
It can be shown along the lines of
\cite{eq} that the system of equations
(\ref{zercur}),(\ref{cons}) expanded near this vacuum solution
properly describes  dynamics
of massless matter fields
on the free field level and beyond.

The main result of this report consists of the observation that
the same system (\ref{zercur}), (\ref{cons}) expanded near another vacuum
solution
describes dynamics of matter fields with an arbitrary mass.
This is a solution with $B_0 =\nu $ where $\nu$ is an
arbitrary constant. For a constant field $B_0$ only the first of the
equations (\ref{cons}) remains non-trivial. Remarkably it turns out to be
possible to find its explicit solution
\be
\label{s0}
{s}_0{}_\ga =z_\ga +\nu (z_\ga -y_\ga )\int_0^1 dt te^{it(z_\ga y^\ga )}K
\ee
(it is not too difficult to check that (\ref{s0}) satisfies (\ref{cons})
 by a direct substitution).
Now let us turn to the equations (\ref{zercur}). The third of these
equations is trivially satisfied. The second one reads
\be
\label{W}
\tilde{W}* s_{0\ga} -s_{0\ga} * W=0\,,\qquad
\ee
where we have taken into account that $ds_{0\ga}=0$.
Eq. (\ref{W}) is
a complicated integral equation.
The key observation however is that it admits
the following two particular solutions: $W_0 =q_\ga$
$(\ga =1,2)$,
\be
q_\ga =y_\ga +\nu K(z_\ga -y_\ga )\int_0^1 dt (1-t) e^{itz_\ga y^\ga }\,.
\ee
Taking into account that $*$-product is associative
it allows us to describe a general solution
of (\ref{W})
as an arbitrary
element $W_0 =\go (q_\ga , K, \psi_{1,2}|x) $ whose arguments
are treated as
some non-commutative elements of the star-product algebra. To make contact
with the previous consideration it remains to check
by the explicit computation that
the elements $q_\ga$ indeed obey
(\ref{modosc}). Thus, the vacuum solution with a constant field
$B_0=\nu$
leads automatically to the deformed oscillator algebra with the
deformed oscillators realized as elements of the tensor product of
two Heisenberg-Weyl algebras.
Finally, it remains to observe that the first of the equations
(\ref{zercur})
reduces to the zero curvature equation which describes the AdS
background space.

Next one can analyze the full system of equations perturbatively by
inserting the expansions of the form: $W=W_0 +W_1 +\ldots$,
$B=B_0 +B_1 +\ldots$ and
$s_{\ga}=s_{0\ga} +s_{1\ga} +\ldots$.
In particular one can derive in the lowest orders that
$B_1 (z,y,K,\psi |x)=
C (q,K,\psi |x),$
$W_1 (z,y,K,\psi |x)=
\go (q,K,\psi |x)+\Delta W_1 (C)$
and $ s_{1\ga} =s_{1\ga} (C)$,
where $s_{1\ga} (C)$ and $\Delta W_1 (C)$ are some functionals
of the field $C$ which remains arbitrary and
has to be identified with generating function
(\ref{maux}). Inserting this back into
(\ref{zercur}) one obtains
in the linearized approximation
the free field
equations for $C$ from the third equation
and the equations of the form $d\go +\go*\wedge\go +O(C^2)=0$
from the first one.
Let us note that in the latter case
the corrections linear in $C$ are compensated by appropriate
field redefinitions so that  the sources for HS
field strengths
(including the gravitational and Yang-Mills ones)
acquire in the lowest order a natural structure of some bilinear currents.

\noindent
\section{Concluding remarks}

\noindent
1. It is shown that global HS symmetries
depend on the mass of
a matter multiplet.

\noindent
2. The full nonlinear HS model admits
continuously degenerate vacua corresponding to
a variety of systems with an arbitrary mass of the matter fields.
The different global HS
symmetries of the linearized matter multiplets are
different stability subgroups of the full HS
symmetry which leave invariant a chosen vacuum solution.

\noindent
3. The model under consideration (eq.(\ref{coeq})) possesses $N=2$
supersymmetry since the deformed oscillator algebra
was shown in \cite{BWV} to contain $osp(2;2)$
generators
$T_{\ga\gb}=\frac{1}{4i}\{q_\ga \,,q_\gb \},$
$Q_\ga =q_\ga $, $S_\ga =q_\ga K$ and
$J=K+\nu$.

\noindent
4. For special values of the parameter $\nu=\frac{2l+1}{2}$,
$\forall l\in {\bf Z}$,
free field equations degenerate in a certain sense and correspond to
 particular gauge systems. The origin of this degeneracy
is that the algebra $Aq(2,\nu )$ admits ideals for these values
of $\nu$ \cite{Q}.
The degenerate
point $\nu=3$ can be shown to
correspond to 3d electrodynamics. Interestingly enough,
the existence of the ideals of  $Aq(2;\nu )$ is related to
the existence of a dual version of the model based on a potential
for a magnetic field.

The research described in this report
 was supported in part by the European Community
Commission under the contract INTAS, Grant No.94-2317 and by the
Russian Foundation for Basic Research, Grant No.96-01-01144.

\end{document}